\documentclass[12pt]{article} 

\usepackage{amssymb} 
\usepackage{amsmath}
\usepackage{amsfonts}

\newcommand{\R}{\mathbb{R}}
\newcommand{\C}{\mathbb{C}}

\newcommand{\be}{\begin{equation}}
\newcommand{\bea}{\begin{eqnarray}}
\newcommand{\eea}{\end{eqnarray}}
\newcommand{\nn}{\nonumber}

\newcommand{\ed}{\end{document}}

\begin{document}

\title{On the Pseudo-Hermiticity of a Class of PT-Symmetric Hamiltonians in One Dimension}
\author{Ali Mostafazadeh\thanks{E-mail address: amostafazadeh@ku.edu.tr}\\ \\
Department of Mathematics, Ko\c{c} University,\\
Rumelifeneri Yolu, 80910 Sariyer, Istanbul, Turkey}
\date{ }
\maketitle

\begin{abstract}
For a given standard Hamiltonian $H=[p-A(x)]^2/(2m)+V(x)$ with arbitrary complex scalar potential $V$ and vector potential $A$, with $x\in\R$, we construct an invertible antilinear operator $\tau$ such that $H$ is $\tau$-anti-pseudo-Hermitian, i.e., $H^\dagger=\tau H\tau^{-1}$. We use this result to give the explicit form of a linear Hermitian invertible operator with respect to which any standard $PT$-symmetric Hamiltonian with a real degree of freedom is pseudo-Hermitian. Our results do not make use of the assumption that $H$ is diagonalizable or that its spectrum is discrete. 
\end{abstract}

\baselineskip=24pt

In a recent series of papers \cite{pt1,pt2,pt3} we have revealed the basic mathematical structure 
responsible for the intriguing spectral properties of certain $PT$-symmetric systems \cite{pt}. The main ingredient leading to the results of \cite{pt1,pt2,pt3} and their ramifications \cite{za1,ply,za2,psusy,solombrino} is the concept of a pseudo-Hermitian operator. By definition,
a linear operator $H$ acting in a Hilbert space is pseudo-Hermitian if there is a linear Hermitian invertible operator $\eta$ satisfying
	\be
	H^\dagger=\eta H\eta^{-1},
	\label{ps}
	\end{equation}
where a dagger stands for the adjoint of the corresponding operator. As we have shown in \cite{pt3}, every diagonalizable Hamiltonian admitting an antilinear symmetry is necessarily pseudo-Hermitian. This applies, in particular, to diagonalizable $PT$-symmetric Hamiltonians. We also explained in \cite{pt1} that most of the $PT$-symmetric Hamiltonians considered in the literature are $P$-pseudo-Hermitian and that there are also $PT$-symmetric Hamiltonians that are not $P$-pseudo-Hermitian. The proof of the equivalence of pseudo-Hermiticity and the presence of antilinear symmetries presented in \cite{pt3} uses the assumption that the Hamiltonian is diagonalizable and has a discrete spectrum. It is based on the fact that every such Hamiltonian $H$ is anti-pseudo-Hermitian, i.e., there is an (essentially unique
\cite{sym,pt3}) antilinear, Hermitian, invertible operator $\tau$ fulfilling
	\be
	H^\dagger=\tau H\tau^{-1}.
	\label{anti}
	\end{equation}
The purpose of this article is to generalize the results of \cite{pt1,pt2,pt3} to the class of standard Hamiltonians: 
	\be
	H'=\frac{1}{2m}\,[p-A(x)]^2+V(x),
	\label{H}
	\end{equation}
with arbitrary complex scalar ($V:\R\to\C$) and vector ($A:\R\to\C$) potentials in one-dimension,
where in the position representation $x\in\R$ and $p=-i\partial_x$. In particular, we wish to address the following two questions without assuming the diagonalizability of the Hamiltonian (equivalently the existence of a complete biorthonormal system of `energy' eigenvectors) or discreteness of its spectrum.
	\begin{itemize}
	\item[1] Is there an antilinear, Hermitian, invertible operator $\tau$ satisfying (\ref{anti}) for such 
	a Hamiltonian? 
	\item[2] Is there a linear, Hermitian, invertible operator $\eta$ satisfying (\ref{ps}) for such a 	$PT$-symmetric Hamiltonian?
	\end{itemize}
In the following $T$ and $P$ denote the time-reversal and parity operators, respectively.

\begin{itemize}
\item[] {\bf Lemma:} Let $\alpha:\R\to\C$ be a complex-valued function. Then $\tau:=T e^{i\alpha(x)}$ is an antilinear, Hermitian, invertible operator.
\item[] {\bf Proof:} $\tau$ is the product of a linear and an antilinear operator, and both of these
operators are invertible. Hence $\tau$ is antilinear and invertible. We can easily establish the 
Hermiticity of $\tau$ using the following simple calculation.
	\[\tau^\dagger=e^{-i\alpha^*(x)}T^\dagger=e^{-i\alpha^*(x)}T=Te^{i\alpha(x)}=\tau,\]
where we have used $T^\dagger=T$ and that for every function $f:\R\to\C$,
	\be
	T f(x) T=f^*(x).
	\label{T1}
	\end{equation}
$\square$
\item[] {\bf Theorem:} Every Hamiltonian of the form (\ref{H}) is anti-pseudo-Hermitian with respect to
	\be
	\tau:=T\,e^{-2i\int_0^x A(x')dx'}.
	\label{tau=}
	\end{equation}
\item[] {\bf Proof:} First recall the identities
	\[e^{-i\alpha(x)}\,p\,e^{i\alpha(x)}=p+\partial_x\alpha(x),~~~~~~~~~
	e^{-i\alpha(x)}\,x\,e^{i\alpha(x)}=x,\]
which are valid for every differentiable function $\alpha:\R\to\C$. Now let 
$\alpha(x):=-2\int_0^x A(x')dx'$ and note that for every function $f:\R\to\C$,
	\be
	T f(p) T=f(-p)^*.
	\label{T2}
	\end{equation}
Then, 
	\bea
	\tau H'\tau^{-1}&=& Te^{i\alpha(x)}\left[\frac{1}{2m}\,[p-A(x)]^2+V(x)\right]\,
	e^{-i\alpha(x)}T\nn\\
	&=&\frac{1}{2m}\,T[p-\partial_x\alpha(x)-A(x)]^2T+V^*(x)\nn\\
	&=&\frac{1}{2m}\,T[p+A(x)]^2T+V^*(x)\nn\\
	&=&\frac{1}{2m}\,[-p+A^*(x)]^2+V^*(x)\nn\\
	&=&\frac{1}{2m}\,[p-A^*(x)]^2+V^*(x)\nn\\
	&=&H^{'\dagger}.
	\label{tau}
	\eea
$\square$
\item[] {\bf Corollary~1:} For every $PT$-symmetric Hamiltonian $H'$ of the form~(\ref{H}),
	\be
	\eta:=e^{2i\int_0^xA^*(x')dx'}P
	\label{eta=}
	\end{equation}
is a linear, Hermitian, invertible operator with respect to which $H'$ is pseudo-Hermitian.
\item[] {\bf Proof:} $\eta$ is the product of two linear invertible operators, so it is linear and invertible.
In order to establish the Hermiticity of $\eta$, we first note that the $PT$-symmetry of $H'$,
	\be
	(PT)H'(PT)=H',
	\label{pt}
	\end{equation}
demands
	\be
	A_r(-x)=A_r(x),~~~~~A_i(-x)=-A_i(x),~~~~~ 
	V_r(-x)=V_r(x),~~~~~V_i(-x)=-V_i(x),
	\label{ri}
	\end{equation}
where the subscripts `r' and `i' respectively denote the real and imaginary parts of the corresponding 
quantity. Now, in view of (\ref{ri}),
	\bea
	\eta^\dagger&=&P e^{-2i\int_0^xA(x')dx'}\nn\\
	&=&e^{-2i\int_0^{-x}A(x')dx'}P\nn\\
	&=&e^{2i\int_0^xA(-x')dx'}P\nn\\
	&=&e^{2\int_0^x[iA_r(-x')-A_i(-x')]dx'}P\nn\\
	&=&e^{2\int_0^x[iA_r(x')+A_i(x')]dx'}P\nn\\
	&=&e^{2i\int_0^x A^*(x')dx'}P\nn\\
	&=&\eta,\nn
	\eea
where we used $P^\dagger=P$ and that for every $f:\R\to\C$,
	\be
	P f(x) P=f(-x).
	\label{p}
	\end{equation}
Next, we recall from \cite{pt3} that the $PT$-symmetry (\ref{pt}) and anti-pseudo-Hermiticity (\ref{tau}) with respect to $\tau$ imply pseudo-Hermiticity with respect to $\tau PT$. But it is an easy exercise to
check that indeed $\tau PT$ with $\tau$ given in (\ref{tau=}) coincides with $\eta$ given in (\ref{eta=}).~~~$\square$
\item[] {\bf Corollary~2:} Every Hermitian Hamiltonian $H'$ of the form~(\ref{H}) admits an antilinear symmetry generated by (\ref{tau=}).
\item[] {\bf Proof:} For a Hermitian Hamiltonian $H'$, Equation~(\ref{tau}) implies $[\tau, H]=0$.~~~$\square$
\end{itemize}

Note that for a Hermitian Hamiltonian $H'$ with a vanishing vector potential $A(x)$, $\tau=T$. This is consistent with the well-known fact that in this case $H'$ is time-reversal invariant. The importance of Corollary~2 is that even when $A(x)\neq 0$, $H'$ admits an antilinear symmetry. In view of the results of \cite{pt3} this is a general property of arbitrary Hermitian Hamiltonians. It may be viewed as a nonrelativistic analog of the $CPT$ theorem.

In summary, we have given an explicit demonstration of the fact that the standard (not-necessarily Hermitian) Hamiltonians (\ref{H}) are anti-pseudo-Hermitian. We used this result to establish the claim that for a standard quantum system with a real degree of freedom $PT$-symmetry is a special case of pseudo-Hermiticity. We wish to emphasize that the results of this paper do not rely on the assumption of the diagonalizability (equivalently the existence of a complete biorthonormal system of `energy' eigenvectors) of the Hamiltonian or discreteness of its spectrum, and in this sense improve those of \cite{pt1,pt2,pt3}. However, we wish to also note that our findings were obtained under the assumption that the configuration space of the system is the real line. This assumption is not valid for all the $PT$-symmetric systems considered in the literature. The next natural step in our study of the relation between pseudo-Hermiticity and $PT$-symmetry is to try to generalize the results of this paper to systems with a complex configuration space.

This project was supported by the Young Researcher Award Program (GEBIP) of the Turkish Academy of Sciences. I would like to thank M.~Znojil for many illuminating discussions.

\ed